\documentstyle[12pt,epsf]{article}

\begin{document}
\def\beq{\begin{equation}}
\def\eeq{\end{equation}}
\def\beqa{\begin{eqnarray}}
\def\eeqa{\end{eqnarray}}
\def\noin{\noindent}
\def\grad{{\bf \nabla}}
\def\bv{{\bf v}}
\def\bB{{\bf B}}
\def\bJ{{\bf J}}
\def\bE{{\bf E}}
\def\D{\Delta}
\def\pa{\partial}
\def\eps{\epsilon_{\alpha\beta}}

\titlepage
\begin{flushright} QMW-PH-97-41
\end{flushright}
\vspace{4ex}

\begin{center} \bf

{\bf\Large Logarithmic Conformal Field Theory  Solutions}\\
{\bf\Large  of Two Dimensional Magnetohydrodynamics}\\

\rm

\vspace{10ex}

Spyros Skoulakis \footnote{email: s.skoulakis@qmw.ac.uk} and
Steven Thomas\footnote{email: s.thomas@qmw.ac.uk}\\
\vspace{10ex}

{\it Department of Physics\\
Queen Mary and Westfield College\\
Mile End Road\\
London E1\\
U.K.}\\

\vspace{14ex}
ABSTRACT
\end{center}
\noindent
We consider the application of logarithmic conformal field theory
in finding solutions to the turbulent phases of 2-dimensional 
models of magnetohydrodynamics. These arise upon dimensional reduction of
standard (infinite conductivity) 3-dimensional magnetohydrodynamics,
after taking 
various simplifying limits. We show that solutions of the corresponding
Hopf equations and higher order integrals of motion can be found within the
solutions of ordinary turbulence proposed by Flohr, based on the tensor
product of the logarithmic extension ${\tilde c}_{6,1} $ of the non-unitary
minimal model $c_{6,1} $. This possibility arises because of the existence
of a continuous hidden symmetry present in the latter models, and the fact that
there appear several distinct dimension -1 and -2  primary fields.

\newpage
\section{ Introduction to Polyakov turbulence and Flohr solution}
The word turbulence is used to describe the irregular and fluctuating motion
that is seen in many cases of gas and liquid flow. We can have a qualitative
description  of the flow by a dimensionless number, the Reynolds number
$Re$, which measures the relative strength of inertial and viscous forces
acting on the fluid. 
Turbulence is associated with flows having very high Reynolds number.
Formally we can take Re $\rightarrow \infty$ and this is the case of the fully
developed turbulent flow. The Navier-Stokes equation describes these flows in 2
and 3 dimensions. One of the  main differences  between  inviscid flows in
each of these dimensions, is the existence of  an infinite number of
conserved quantities in 2-dimensions, and it is these and
their corresponding conserved fluxes that make two dimensional turbulence
particularly interesting. Indeed there has been a growing body of research
on the subject since the classic work of Kraichnan [1]. 

A fresh approach to the problem of understanding turbulent flows
in $d=2 $ was presented by Polyakov [2,3], who, 
motivated by the success of Conformal Field
Theory (CFT) in describing 2D critical phenomena, suggested that the
methods of CFT can help in this task . The central point in the analysis
was the Hopf equation, written here in terms of the vorticity 
$ \omega = \epsilon_{\alpha\beta}\pa_{\alpha} v_{\beta},~~~
\alpha,\beta=1,2 $
\beq\label{eq:1.1}
\sum_{p=1}^{n} \langle \,  \omega(x_{1}) \cdots \partial_{t} \omega(x_{p}) \cdots \omega
(x_{n}) \, \rangle = 0 
\eeq
where the time derivative of $\omega$ appearing in eq(\ref{eq:1.1}) is given
in terms of the Navier -Stokes equations
\beq\label{eq:1.2}
\partial_{t} \omega +\epsilon_{\alpha \beta}\partial_{\alpha} \psi 
 \partial_{\beta} \omega = \nu \partial^{\alpha}
\partial_{\alpha} \omega \quad 
\eeq
where, assuming incompressibility, the velocity is expressed in terms  of the stream function $\psi $ 
as  $v_{\alpha} = \epsilon_{\alpha \beta} \partial_{\beta}
\psi $
    
The Hopf equations  express  $n$-point function in terms of the  $n+1$ point
functions. They can't be solved directly and the only way so far was to
take ad-hoc approximations and try to find solutions. In contrast, with the
Conformal approach we try to solve the Hopf equation exactly by the ansatz
that the flow has conformal symmetry in the situation where it is fully developed.   

The assumption is that there exists a range of scales such that the affect
of the 
viscocity  term is very small - the so called inertial range. So 
 we insert $ \omega = - \partial^2 \psi  $ in the Hopf equation, postulating that
 $\psi$ is a primary operator of the CFT we are looking for. Because of
 this, one is naturally lead to  
introduce point splitting regularization in order to evaluate products of fields defined 
at the same point. Using the  operator product expansion (OPE) for $\psi $
with itself
\beq\label{eq:1.3}
[\psi ] \times [\psi] =[\phi ]+ ....
\eeq
where $\phi$ is the  conformal field with the minimal dimension appearing in the
OPE,  and  $+.... $ indicates next to leading order terms, 
the Navier-Stokes equations  imply that 
\beq\label{eq:1.3b}
\dot{\omega}= - |a|^{2(\Delta_{\phi}-2\Delta_{\psi})}{\cal L}\phi,\qquad
{\cal
L}=[L_{-2}\bar{L}_{-1}^{2}-\bar{L}_{-2}L_{-1}^{2}]
\eeq
appears inside the correlators in the Hopf equation, 
where $a$ is the u.v. cutoff associated with point splitting. In principle, 
a solution to the Hopf equations can be found if the operator on the right
hand side in eq(\ref{eq:1.3b}) is vanishing.
However, this requirement by itself is 
not restrictive enough to allow us to find explicit solutions.
To place further restrictions, one has to take into account the spectrum 
associated with turbulence.
 One of the predictions of the conformal approach  is that the  $k$-space 
kinetic energy density $E(k) $ (defined by  $ \int dk E(k) =  \langle \,
v_{\alpha}(k) v_{\alpha}(-k) \, \rangle $  ) implies a power law spectrum with 
$ E \sim  k^{4 \Delta_{\psi } + 1} $. In the classical
turbulence studies, the value of the exponent of $k$ in the above relation
was one of  the first 
things to be studied. There are different opinions for its value but the work of
Kraichnan [1] indicated that in the case of the inverse cascade scenario,
 (similar
to the Kolmogorov one in  $d=3$),  $E\sim k^{-3} $
which is based on the constancy of the
enstropy flux. So it is natural to postulate this cascade scenario in order to
increase our restrictions on the possible conformal field theory
solutions. In this  case we have

\beq\label{eq:1.4} 
\langle \, \dot{\omega} (z, \bar{z}) \, \omega(0) \, \rangle \sim {\rm const}
 \quad \Rightarrow \quad
 \Delta_{\phi} +\Delta_{\psi} +3 =0
\eeq

Trying to find solutions of these constraints, Polyakov considered the
 possibility 
$2\Delta_{\psi} < \Delta_{\phi}$, which `trivially' satisfies
the Hopf equations in $|a| \rightarrow 0 $ limit, 
and found solutions within the non-unitary minimal models.
 In fact hundreds of solutions can be found in this way
[4-9], which is perhaps an indication that this approach is not 
encapsulating all the physical aspects of turbulence we would like.
Worse still, Falkovich and Hanany [10], showed that a solution that
 has the Kraichnan
spectrum can not be found within the minimal models. Because 
of the  constancy of all the higher order fluxes,  they found that 
$2\Delta_{\psi} = \Delta_{\phi} =-2$ and this case is not found in the minimal
models. Also in the minimal model case the actual prediction is that all
fluxes vanish so the situation is made more difficult.
Such difficulties indicate perhaps that the `usual' Conformal approach might
describe certain kinds of  flows,  but not the turbulent ones.

In an attempt to directly deal with these issues, Flohr [11] has
 suggested a solution that does not suffer  from  these
problems and is based on Logarithmic Conformal Theories (LCFT) [12 -16].
This  ability to avoid the aforementioned difficulties and yet still remain
within the conformal approach is clearly due to the novel 
characteristics of LCFT's. 

One such  important characteristic, is the existence of pairs of primary 
fields $C(x), D(x) $ in the spectrum having the same dimension and with
respect to which the Virasoro operator $L_0 $ is non-diagonal i.e.
$L_0 | C \rangle  = \Delta_C  | C \rangle , \quad  L_0 | D \rangle = \Delta_D | D \rangle  +  \vert C \rangle $.  

The two  point functions of the logarithmic pairs $C(x), D(x) $
have the following form (see e.g. [13] )

\beqa\label{eq:1.5}
\langle C(x) D(y) \rangle  &=& \frac{d}{(x-y{)}^{2 \Delta_C} } \cr
&&\cr
\langle D(x) D(y) \rangle & = & \frac{d}{(x-y{)}^{2 \Delta_C} } 
( - 2 e\,  {\rm ln} (x-y) + d ) \cr
&&\cr
\langle D(x) D(y) \rangle & =& 0
\eeqa
where $d, e $ are constants.
As pointed out in [13], the first correlation in eq(\ref{eq:1.5}) together
with the invariance of the correlator under exchange of $C(x) $ and $D(x) $
implies a strong constraint on the allowed values of the dimensions
$\Delta_C$, namely that it be integer.\footnote{See also [26] for
  additional clarification of this point}

 This implies that the logarithmic
operators $C(x)$ (or rather their chiral components) are associated with a
hidden continuous symmetry, although unlike the more usual current algebras,
the Schwinger term is vanishing.

This  hidden continuous  symmetry associated with LCFT's is an
important ingredient in Flohr's solution of the Hopf equations
(\ref{eq:1.1}). 

In searching for possible non-unitary LCFT solutions to turbulence, (the
non-unitarity is a consequence of the flux constraint (\ref{eq:1.4})), 
the logarithmic extensions of the minimal model CFT's is perhaps the
first place to start since a certain amount of knowledge already exists
about their properties [16]. In fact it is the ${\tilde{ c}}_{p,1} $ models that have
been studied most. As we have seen above, the Kraichnan spectrum implies
that there should be $\Delta = -1 $ and $-2 $ states in the spectrum of our
LCFT. This is not possible in a single copy of ${\tilde{c}}_{p,1} $ but as shown by
Flohr [11], since ${\tilde{c}}_{6,1} $ does contain a $\Delta = -1 $ field, the tensor
product ${\tilde{c}}_{6,1} \otimes    {\tilde {c}}_{6,1} $ can contain both. The additional
advantage of the tensor product model is that it naturally contains 
Polyakov's Bose condensate [3] within the solution.   
 
${\tilde{c}}_{6,1} $ has  17 primary fields, 5 of them are "logarithmic". The
dimensions of the fields are $ \Delta_{1,r}=\frac{1}{4p}((p-r)^{2}-(p-1)^{2})$
with $p=6$.

Because of the tensor product there is a degeneracy in the way we can
define the
states with dimension $-1, -2$ . In fact this will prove to be a nice
feature of the solution, which we shall exploit later on when considering
the MHD case. 

We define the  $ \Delta  = -1 $  and  $ \Delta = -2 $   fields, 
$\psi_i , \phi_{j'} $  as follows (with 
$\Phi_{(\D \mid  \D')} \equiv (\Phi_{\D}\otimes \Phi_{\D'})(z)$, and $\Phi_{\D} (z) $ a primary field in 
  $ \tilde{c}_{6,1} $ )
\beqa\label{eq:1.6}
\psi_1 & = & \frac{1}{2} ( \Phi_{(-1 \mid -\tilde{0})} + \Phi_{(-\tilde{0}
\mid -1)} ), \quad  
\psi_2  =  \frac{1}{2} ( \Phi_{(-\tilde{1} \mid -\tilde{0})} + \Phi_{(-\tilde{0}
\mid -\tilde{1})} ) \cr 
\phi_{1}& = & \Phi_{(-1 \mid  -1 )}, \quad  \phi_{2} =\frac{1}{2} ( \Phi_{(-1 \mid
-\tilde{1})} + \Phi_{(-\tilde{1} \mid -1) }),\quad  \phi_{3}=
\Phi_{(-\tilde{1} \mid -\tilde{1} )} 
\eeqa
It is easy to see that $\psi_i \times \psi_j = \sum_{k'}C^{k'}_{ij}
\phi_{k'} + ...  \, , i, j = 1,2 $ and $k' = 1,2,3 $, where the OPE coefficients $C_{ijk'} $
can be obtained from the fusion rules of $\tilde{c}_{6,1} $ LCFT given in 
[16].

We expect the physical fields to be given by linear combinations of the 
fields $\psi_i $ and $\phi_{j'} $, i.e. $\psi_{ph} = \sum_i \alpha_i
\psi_i $ and $\phi_{ph} = \sum_{j'} \beta_{j'} \phi_{j'} $ with 
\beq\label{eq:1.7}
\psi_{ph} \times \psi_{ph} \, = \,  \phi_{ph} + .....
\eeq
In order that the physical fields $\psi_{ph} $ satisfy eq(\ref{eq:1.7}),
the coefficients $\alpha_i $ and $\beta_{j'} $ must satisfy 
$\sum_{ij} \alpha_i \alpha_j C_{ijk'} = \beta_{k'} $ which essentially
determines the coefficients $\beta_{k'} $ once the  $\alpha_i $ are known.

From the general form of $\psi_{ph} $ one can determine that its $k$-space
two-point function has the structure
\beq\label{eq:1.8}
  \langle \psi_{ph} (k) \psi_{ph} (-k) \rangle   =  \frac{a_{0}}{ \mid k
    \mid^{2-4 \D_{\psi}}}
+\frac{a_{1}}{L^{2 \D_{\psi}}  } \delta(k)  + ....
\eeq
with the coefficients $a_0 \sim \langle I \rangle $ and 
$a_1 \sim \langle \phi_{ph} \rangle $. 
A notable feature as pointed out in [11] , is the presence of the 
so called Polyakov condensate term (proportional to $\delta (k) $)
which is naturally contained within the solution.  

Flohr considered a more general expansion of the physical fields 
$\psi_{ph} $ and $\phi_{ph} $ involving non-diagonal
combinations of left and right-moving states. This leads to non-zero
spin contributions to $\psi_{ph} $ and $\phi_{ph} $ that may naturally
encode the effect of stirring forces in the fluid. The general expansions
are of the form

\beqa\label{eq:1.9}
{\hat{\psi}}_{ph}(z , \bar{z} ) & = & \psi_{ph} (z) \times  \psi_{ph} ( \bar{z}
) + \sum_{\D_U + {\bar{\D}}_V = -2,\, \D_U + {\bar{\D}}_V \epsilon Z } 
U(z) \times V(\bar{z} ) \cr
&&\cr
{\hat{\phi}}_{ph}(z , \bar{z} ) & = & \phi_{ph} (z) \times  \phi_{ph} ( \bar{z}
) + \sum_{\D_{U'} + {\bar{\D}}_{V'} = -4,\, \D_{U'} + {\bar{\D}}_{V'} \epsilon Z } 
U'(z) \times V'(\bar{z} ) 
\eeqa
 where the integer dimension fields $U(z), V(\bar{z}), U'(z), V'(\bar{z} ) $
are those occuring in ${\tilde{c}}_{6,1} \otimes {\tilde{c}}_{6,1} $. This leads to the following expression for ${\hat{\psi}}_{ph} $

\beqa\label{eq:1.10}
{\hat{\psi}}_{ph} (z, \bar{z} )& = & 
a_{1}\Phi_{(-1 \mid 0)}(z) \otimes \bar{\Phi}_{(-1 \mid 0)}
(\bar{z}) + a_{2}\Phi_{(0 \mid -1)}(z) \otimes \bar{\Phi}_{(0 \mid -1)}
(\bar{z})  \cr
& + & a_{3} \Phi_{(-1 \mid \tilde{0})}(z) \otimes \bar{\Phi}_{(-1 \mid
\tilde{0})} (\bar{z}) +a_{4}\Phi_{(\tilde{0} \mid -1)}(z) \otimes
\bar{\Phi}_{(\tilde{0} \mid -1)} (\bar{z})  \cr
&+&a_{5}\Phi_{(-1\mid -1)}(z) \otimes
\bar{\Phi}_{(0 \mid 0)} (\bar{z})+ a_{6} \Phi_{(0 \mid 0)}(z) \otimes \bar{\Phi}_{(-1\mid -1)} (\bar{z}) \cr
&+& a_{7}  \Phi_{(0 \mid 0)}(z) \otimes \bar{\Phi}_{
(\tilde{0} \mid \tilde{0} )(\bar{z})}  + \cdots
\eeqa
There is a similar expansion of ${\hat{\phi}}_{ph} (z, \bar{z} ) $.
The non-diagonal terms in eq(\ref{eq:1.10}) induce further terms in the 
two-point function eq(\ref{eq:1.8}) that are proportional to 
the expectation values of chiral fields, and involve the derivative of 
$k$-space delata functions [11].
A crucial aspect of the model is the existance of a hidden continuous 
 symmetry [13] associated with the integer dimension field 
$\Psi_{\D_{1,5}} $. In fact this
guarantees that it is a solution to the Hopf equations.
The operator $\Omega (z) = L_0 \bar{L}_0
[L_{-2}\bar{L}_{-1}^{2}-\bar{L}_{-2}L_{-1}^{2}] \phi_{ph} (z) $ generates a continuous
symmetry of the LCFT  and 
satisfies the condition \footnote{As argued in [11],  
The presence of $L_0 \bar{L}_0 $ in $\Omega (z) $ is needed to 
ensure that only the $\Psi_{\D_{1,5}} $ part of $\phi_{ph} $ enters the
Hopf equations. This follows from certain properties of correlation functions 
of LCFT discussed in [15] }  

\beq 
\langle  \Omega  (z) \Phi_{1}(z_{1}) \Phi_{2}(z_{2}).... \Phi_{n}(z_{n})
\rangle = 0
\eeq

In ending this section, 
it is perhaps worth mentioning the reason 
 one finds non-unitary CFT in connection with turbulence,
 may in some sense be
expected, since the turbulent state is not an equilibrium state. Even if the 
conformal solutions don't describe the turbulent state exactly, we 
might at least expect them
to describe flows out of equilibrium.

 Motivated by the new possibilities that LCFT has brought to 
conformal turbulence, in the next section
 we want to extend Flohr's construction and 
look for solutions of the generalized  turbulent 
MHD systems  in $d=2$ that arise naturally when dimensionally reducing
standard ideal MHD in $d=3$. 

\section{ Logarithmic solutions to MHD turbulence}

The long wavelength limit of 3-dimensional plasmas are well approximated 
by the equations of Magnetohydrodynamics (MHD), which in the limit of 
low frequencies, low temperature and incompressibility take the simplified
 form [17]
\beqa\label{eq:2.1}
\grad\cdot\bv & = & 0 \nonumber \\
\grad\cdot\bB & = & 0 \nonumber \\
\pa_{t}\bv + \bv\cdot\grad\bv & = & -\frac{1}{\rho_{m}}\grad P +
\frac{1}{c\rho_{m}}\bJ
\times\bB \\
\grad\times(\bv\times\bB) & = & \pa_{t}\bB \nonumber \\
\grad\times\bB & = & \frac{4\pi}{c} \bJ \nonumber
\eeqa

\noin
Here $\bv , \bB , \bJ $ are the velocity field, magnetic field and 
electric current respectively. $P$ and $\rho_{m} $ are the  pressure
 and  mass density.
After simple manipulations, we can recast these equations into a more
symmetric form 
\beqa\label{eq:2.2}
\grad\cdot\bv & = & 0 \nonumber \\
\grad\cdot\bB & = & 0 \nonumber \\
\grad\times(\pa_{t}\bv + \bv\cdot\grad\bv) & = &
\frac{1}{4\pi\rho_{m}}\grad\times
(\bB\cdot\grad\bB) \\
(\bB\cdot\grad)\bv - (\bv\cdot\grad)\bB & = & \pa_{t}\bB \nonumber
\eeqa

In order to obtain an effectively two-dimensional theory from 
eq(\ref{eq:2.2}), one simple proposal is to dimensionally reduce the model  
by requiring that all fields be independent of the $z$-coordinate, i.e.
$ \pa_{3}\bv = 0 $ and  $\pa_{3}\bB = 0 $.
It follows that the first two components of $\bv$ and
$\bB$ can be written as
\beq\label{eq:2.3}
B_{\alpha} = \epsilon_{\alpha\beta}\pa_{\beta}A,~~~v_{\alpha} = \epsilon_
{\alpha\beta}\pa_{\beta}\psi,~~~~~ \alpha,\beta=1,2.
\eeq

\noin
One can  define, in analogy with vorticity $\omega$, the ``magnetic vorticity"
\beq\label{eq:2.4}
\Omega\equiv \epsilon_{\alpha\beta}\pa_{\alpha}B_{\beta}=-\pa_{\alpha}
\pa_{\alpha}A,
\eeq
and the two-dimensional operator
\beq\label{eq:2.5}
{\cal A}\equiv \eps\pa_{\beta}A\pa_{\alpha}
\eeq
after which the 2-D MHD equations take the form [18]:
\beqa\label{eq:2.6}
\dot{\omega} + \eps\pa_{\beta}\psi\pa_{\alpha}\omega & = &
\frac{1}{4\pi\rho_{m}}{\cal A}\Omega \nonumber \\
\dot{A} + \eps\pa_{\beta}\psi\pa_{\alpha}A & = & 0 \nonumber \\
\dot{V} + \eps\pa_{\beta}\psi\pa_{\alpha}V & = & \frac{1}{4\pi\rho_{m}}{\cal
A}B
\\
\dot{B} + \eps\pa_{\beta}\psi\pa_{\alpha}B & = & {\cal A}V
\nonumber
\eeqa

Solutions to eqs(\ref{eq:2.6}) have been studied in the context of 
minimal model CFT, in two simplifying
cases, namely parallel flow conditions where $B = V = 0$ [18], 
and perpendicular flow for which $ A = V = 0 $ [19]. 
In both cases two of the four equations in (\ref{eq:2.6}) are automatically
satisfied and the remaining equations to be studied are:

\beqa\label{eq:2.7}
\dot{\omega} + \eps\pa_{\beta}\psi\pa_{\alpha}\omega & = &
\frac{1}{4\pi\rho_{m}}{\cal A}\Omega \nonumber \\
\dot{A} + \eps\pa_{\beta}\psi\pa_{\alpha}A & = & 0  
\eeqa
for parallel flow and 

\beqa\label{eq:2.8}
\dot{\omega} + \eps\pa_{\beta}\psi\pa_{\alpha}\omega & = & 0  \cr
\dot{B} + \eps\pa_{\beta}\psi\pa_{\alpha}B & = & 0 
\eeqa
for perpendicular flow.

We want to discuss the behaviour of various fluxes when one considers 
dissapative forces. In the case of ordinary fluids this means
reintroducing the viscosity $\nu $. In the MHD case there is an additional
dissapation parameter namely the conductivity $\sigma $. Keeping 
$\nu = 0 $ for the present but $\sigma $ finite, we find for  
the equations of motion of perpendicular flow [19]  

\beqa\label{eq:2.9}
\dot{\omega} + \eps\pa_{\beta}\psi\pa_{\alpha}\omega & = & - \frac{\rho_{c}}{\rho_{m}}
 (\frac{c}{4\pi\sigma}\pa_{\alpha}\pa_{\alpha}B-\frac{1}{c}[\eps\pa_{\beta}\psi
 \pa_{\alpha}B ]) \\
\dot{B} + \eps\pa_{\beta}\psi\pa_{\alpha}B & = & \frac{c^2}{4\pi\sigma}\pa_{\alpha}\pa_{\alpha}B  \nonumber 
\eeqa

Finite viscosity $\nu $ would add the standard dissapative term to the rhs
of the $\omega $ equation of motion but leave that of $B$ unaffected.
Hence viscosity and conductivity are responsible for the decay of 
higher point enstrophy and magnetic fluxes, $H_n (k,t) $ and 
${\cal B}_n (k, t) $ where

\beq\label{eq:2.10}
 \int d k H_n (k, t)  =  \int d^2 x  \langle \, \omega^n (x)\, \rangle , \qquad
\int dk {\cal B}_n (k , t) =  \int d^2 x  \langle \,  B^n (x)  \rangle     
\eeq

For infinite conductivity but non-vanishing $\nu $, 
${\cal B}_n $ is time independent  (so  $B$-fluxes are conserved) but 
$H_n $ satisfies 

\beq\label{eq:2.11}
\dot{H}_n \sim \nu \int^{a^{-1}}_{L^-1} k^2 h_n (k,t) d k 
\eeq
where we have introduced the I.R. cutoff $L$.
Dimensional analysis implies that the viscosity $\nu $ and ultraviolet 
cutoff $a$ are related via $\nu \sim a^{2 \D_{\psi} - 2 \D_{\phi} } $.
A further analysis on the viscosity dependence of $H_n (k) $ implies [5] 

\beq\label{eq:2.12}
\dot{H}_n \sim \nu^{[ (n-1)(\D_{\psi} +1) + \D_{\phi} +2 ] / (\D_{\phi} -
\D_{\psi} ) } 
\eeq

Demanding that the higher order fluxes $H_n $ have the same scaling 
properties independent of $n$ determines the scaling dimensions of 
both $\phi $ and $\psi $ to be $\D_{\psi} = -1 , \D_{\phi} = -2 $ 
and immediately gives rise to the Kraichnan form for the energy spectrum
$E(k) \sim k^{-3} $ [5]. 

In the present case of MHD we can also imagine considering the situation
where we have inviscid flow but with $\sigma $ finite. In this case we can
perform a similar dimensional analysis as before. The relation between 
$\sigma $ and $a$ can be determined from the $B$ equation of motion (20) 
to be $\sigma^{-1}  \sim a^{( 2 \D_B - 2 \D_{\chi'} -2 ) } $,
where $\chi' $ is defined to be the minimal dimension field in the OPE 
$ B \times \Psi $. 
The fluxes $H_n $ and ${\cal B}_n $ then satisfy
\beqa\label{eq:2.13}
\dot{H}_n & \sim & \frac{\rho_e}{\rho_m}  \sigma^{-1} \int d^2 x \langle \,
\omega^{n-1}
\partial^2 B \, \rangle + ... \cr 
\dot{\cal{B}} &\sim &  \sigma^{-1} \int d^2 x \langle \,  B^{n-1}
\partial^2 B \, \rangle  
\eeqa

It follows that the operators $\dot{\omega} $ and $\dot{B} $ have the same
dimension and hence $\D_{\chi' } = \D_{\phi} = -2 $ for consistency. The $\sigma $
dependence of  $ {\cal B}_n $ can be obtained as 

\beq\label{eq:2.14}
{\cal B}_n \sim \sigma^{[ (n-2) \D_{B} +\D_{\chi'}+ 2 ] / ( \D_{B} - \D_{\chi'} )}
\eeq

Hence it follows that the scaling behaviour of ${\cal B}_n$ is independent
of $ n$ if $ \D_{B} = 0 $ and $\D_{\chi'} = -2 $, and we see  the condition on $\D_{\chi' } $ 
agrees  with the one found above by consistency.

In studying the possible solutions of the inviscid ideal MHD equations,
in the perpendicular flow regime
eq(\ref{eq:2.8}), we have to identify the dimension zero field $B$
with certain primary operators. We want to look for solutions of the 
resulting  Hopf equations similar to the logarithmic conformal
field theory solutions considered by Flohr in the case of normal fluids.
An obvious motivation for this is that the LCFT $ {\tilde{c}}_{6,1} \otimes 
{\tilde{c}}_{6,1} $ considered in [11] solves the Hopf equation  involving $\omega $,
which is one of the equations in (\ref{eq:2.8}). The hope is (and we shall
see that this is indeed possible) that one can identify the $ B$ field 
with primary fields within this particular LCFT in such a way as the 
 Hopf equations associated with the second equation in (\ref{eq:2.8})
can be satisfied. Even though this is in the context of perpendicular flow,
one should expect that similar ideas would apply to the parallel case as well.

Indeed, considering  the equation of parallel flow, eq(\ref{eq:2.7}), one obtains
 after performing the regularization and  OPE's,
\beqa\label{eq:2.17}
\dot{\omega}&= &|a|^{2(\Delta_{\varphi}-2\Delta_{A})}{\cal L}\varphi - |a|^{2
(\Delta_{\phi}-2\Delta_{\psi})}{\cal L}\phi = 0\cr
\dot{A} & = & |a|^{2(\Delta_{\chi}-\Delta_{A}- \Delta_{\psi} )} {\cal L}\chi = 0
\eeqa
where in eq(\ref{eq:2.17}), ${\cal
L}=[L_{-2}\bar{L}_{-1}^{2}-\bar{L}_{-2}L_{-1}^{2}]$
and $A \times A=[\varphi ]+\cdots, \psi \times \psi=[\phi]+\cdots$
and $ A \times \psi  = [\chi] + \cdots $. Here
$\varphi, \phi $ and $ \chi $ are minimal dimension fields.

Perpendicular flow corresponds to the equations

\beqa\label{eq:2.18}
\dot{\omega} &=&- |a|^{2(\Delta_{\phi}-2\Delta_{\psi})}{\cal L}\phi = 0 \cr
\dot{B} &=&- |a|^{2(\Delta_{\chi' }-\Delta_{\psi}-\Delta_{B} )}{\cal
 L}\chi' = 0 
\eeqa
where $\chi' $ is the minimal dimension field in the OPE $ B \times \psi = 
[ \chi' ] +  \cdots $.

Now we have seen previously that flux conditions dictate that $\Delta_B = 0
$. Also one finds that on dimensional grounds $\Delta_A = 0 $. Thus both
these fields have to be constructed from  $ \Delta = 0 $ fields in the 
$ {\tilde{c}}_{6,1} \times {\tilde{ c}}_{6,1} $ LCFT. Just as in the case of the $ \Delta = -1,
-2 $ states, there are several possible primary fields of dimension zero,
which we call $b_{i'} , i' = 1,2,3 $,  where 

\beqa\label{eq:2.19}
b_1 &=& \Phi_{(\tilde{0}| \tilde{0} )}, \qquad b_2 = \frac{1}{2} ( 
\Phi_{(1|-1)} + \Phi_{(-1|1)} ) \cr
b_3 &=& \frac{1}{2} ( \Phi_{(1|-\tilde{1})} + \Phi_{(-\tilde{1}|1)}  )
\eeqa 

(There are other possibilities  involving tensor products containing
the identity operator $I$ which we do not include as they are in some sense
trivial).

As in the case of ordinary turbulence, we expect that the physical 
components of the magnetic field $B$ and $A$ will be given by some linear
combination of the $b_i $ fields. Under these circumstances, in verifying 
either of the two sets of Hopf equations following from eqs(\ref{eq:2.17},
\ref{eq:2.18}) we need to calculate the OPE's  $b_{i'} \times b_{j'}
$ and $b_{i'} \times \psi_j $. In order to obtain these
expansions we need some further  fusion rules of $\tilde{c}_{6,1} $ beyond
those considered in [11]. Specifically we need those involving the primary
 field $[1] $ with the set $[\tilde{0}], [-1], [-\tilde{1}], [1] $, which 
can be obtained from the general rules presented in reference [15],

\beqa\label{eq:2.20}
[1] \times [\tilde{0}]& =& [-\frac{2}{3}] + [-1] +[-\tilde{1}] +
[-\tilde{\frac{2}{3}}] +[\tilde{0} ] + [1] +[\frac{7}{3}] +[4]\cr
[1] \times [-1]& =& [-\tilde{\frac{2}{3}}] +[ \tilde{0}]+[1] +[\frac{7}{3}]
+[4] \cr
[1] \times [-\tilde{1}]& =& [-\tilde{1}] + [ -\tilde{\frac{2}{3}}] +
[\tilde{0}] + [1] + [ \frac{7}{3}] +[4]\cr
[1] \times [1]& =& [  \tilde{0}] + [-\frac{2}{3}] + [-1] +
 [-\tilde{\frac{2}{3}}] + [1] +[\frac{7}{3}] +[4] 
\eeqa                   
where the right hand side of the OPE's in eq(\ref{eq:2.20}) is to be
understood  modulo multiplicities and subject to the linear relation
amongst  certain conformal families  as explained in detail in [16].
 Using these
relations and those defined earlier in section 1, one can obtain 

\beqa\label{eq:2.21}
b_{1} \times \psi_{1}& =& 4 \phi_{2}  \cr
b_{1} \times \psi_{2} & = & 2 \phi_{1}\cr
b_{2} \times \psi_{1} & = &\frac{1}{2} \phi_{1} +\phi_{2} +\frac{1}{2} \phi_{3}\cr
b_{2} \times \psi_{2} & = &  \phi_{2} + \phi_{1}\cr
b_{3} \times \psi_{1} & = & \frac{1}{2} \phi_{1} + \frac{1}{2} \phi_{2}\cr
b_{3} \times \psi_{2} & = & \phi_{2} + \frac{1}{2} \phi_{1}\cr
b_{1} \times b_{1} & = &\phi_{1}\cr
b_{1} \times b_{2} & = & \frac{1}{2} \phi_{1} + \phi_{2} + \frac{1}{2} \phi_{3}\cr
b_{1} \times b_{3} & = & \phi_{1} + \phi_{2}\cr
b_{2} \times b_{3} & = & \frac{1}{4} \phi_{1} + \frac{1}{2}\phi_{2}\cr
b_{3} \times b_{3} & = & \frac{1}{2} \phi_{1} + \frac{1}{2} \phi_{3} +
\frac{1}{2} \phi_2
\eeqa 

An important feature of the OPE in eq(\ref{eq:2.21}) is that the leading
terms on the right hand side (i.e. those with smallest dimensions) are
those of $[-1] $ and $[-\tilde{1}] $, except in the OPE $[1] \times [-1] $ 
where the lowest dimension is $-2/3$. Despite this exception one finds that 
the leading order representations on the rhs of eq(\ref{eq:2.20}) are 
just given by linear combinations of the $\phi_i $ fields defined earlier. 

In the case of parallel flow the solution includes a field with dimension
 $-1 $
($\psi$), one field with dimension $0$ ($A $) and 3 fields with dimension
$ -2 $,
($\phi,\chi, \varphi$).
In the case of perpendicular flow we have again $\psi$,   $B$ with
dimension
$0$
and $\chi' $ with dimension  $-2 $.
We see that the solution accommodates both cases because of the freedom we have
in the definition of the physical fields. We define the physical  fields 
$\phi_{ph}, \psi_{ph}$
as in the previous section (for simplicity we do not consider 
non-diagonal combinations at this point), and
 $A_{ph}= \sum_{i'} \gamma_{i'} b_{i'},\quad B_{ph}= 
\sum_{i'} \delta_{i'} b_{i'},\quad \chi_{ph}= \sum_{i'} \zeta_{i'} \phi_{i'} , \quad
 \varphi_{ph} =
\sum_{i'} \eta_{i'} \phi_{i'}$ and $ \acute{\chi_{ph}} = \sum_{i'} \theta_{i'} \phi_{i'}$.
The additional fusion rules involving $b_{i'} $ that we need are  
$ b_{i'} \times b_{j'}= \sum_{k'}E_{i'j'}^{k'} \phi_{k'} , \quad 
 b_{i'} \times \psi_{j} = \sum_{k'}D_{i'j}^{k'} \phi_{k'}$

Previously, we discussed the constraint on the coefficients
$\alpha_i , \beta_{j'} $ originating from condition that
 the OPE of $\psi_{ph} $
 with itself give $\phi_{ph} $ as a minimal dimension field.
This requirement is an integral part of the conformal approach to solving
the Hopf equations. In the present case  we have similar conditions 
which must be imposed on the coefficients occuring in the 
definitions of $A_{ph}, .. \chi'_{ph} $. Specifically we have  
\beqa\label{eq:2.22}
A_{ph} \times A_{ph} &=& \varphi_{ph}  \rightarrow  \sum_{i'j'}\gamma_{i'}
\gamma_{j'}E_{i'j'}^{k'}= \eta_{k'} \cr
A_{ph} \times \psi_{ph}&=& \chi_{ph} \rightarrow 
\sum_{i' j} \gamma_{i'}\alpha_{j} D_{ij}^{k'}
= \zeta_{k'}\cr
B_{ph} \times \psi_{ph}& =& {\chi '}_{ph} \rightarrow
\sum_{i'j} \delta_{i'}
\alpha_{j} D_{i'j}^{k'} = \theta_{k'}
\eeqa
So if we know $\alpha_{i}$ and $\gamma_{i'}$ we can find the rest of the
coefficients in the case of the parallel flow. The same holds for the
perpendicular case if we know $\alpha_{i}$ and $ \delta_{i'}$.
 In fact we expect that a change in  these
parameters is related to  changing the boundary conditions 
that leave the spectrum invariant and so represent a kind of universality.
 These are the free parameters of the solution.

At this stage we wish to make a few comments concerning solutions 
presented.
It is apparent from Flohr's work [11] and the results presented here,
 that  taking copies of a conformal model provides a  way to incorporate the 
infrared condensate. If the copies are different then we have an asymmetry as
can be seen from the following.
Take as another  possible solution  $ \tilde{c}_{6,1} \otimes c_{6,1} $. Then we have 
two $\D = -2 $ and two $\D= -1 $ fields, $ \phi_{1}= \Psi_{(-1 \mid -1)},\quad 
 \phi_{2}= \Psi_{
(-\tilde{1} \mid -1)}, \quad  \psi_{1}= \Psi_{( -\tilde{1} \mid 0) },\quad
  \psi_{2}= \Psi_{(
\tilde{0} \mid -1) }$. However there is  only one  $\D = 0$ field, namely  $b=
\Psi_{(1 \mid -1)}$. It follows that
$ b\times b= \phi_{1}$ and so this naturally provides another solution 
of the case of perpendicular flow only. There may be other asymmetric
tensor product solutions of this type.

\section {Conclusions}
In this paper we have considered the possibility of describing solutions of
$d=2 $ ideal MHD turbulence within the framework of Flohr's logarithmic
field theory solution in the pure fluid case. We have seen that the 
$\tilde{c}_{6,1}\otimes \tilde{c}_{6,1} $ theory has sufficient degeneracy
of dimension $0, -1 $ and $-2  $ fields to be able to describe the
additional fields one finds in the MHD case. Moreover the solutions of the
corresponding Hopf equations was seen to be a consequence of the same hidden
continuous symmetry that was exploited in the ordinary fluid case.
This picture provides an interesting viewpoint of turbulent 2-d fluids and
plasmas as corresponding to different regions of parameter space 
within  a single model.

It would also be interesting to clarify further the exact relationship between
the values of the real parameters entering the definitions of the physical
fields $\psi_{ph}, \phi_{ph}, A_{ph} $ and $B_{ph} $ given earlier, and 
boundary conditions that would be appropriate in a given physical
situation. This question was already touched upon in [11]. Also one can 
imagine that our results should be extendible to the case of ``dyonic'' or 
duality invariant MHD [20 ], whose minimal model conformal
solutions have been investigated in [21].   

Finally, there has been some  interest in the past, in studying 
perturbations of CFT solutions to turbulence which break the conformal
symmetry [22-24]. The application of such perturbations in 
LCFT's has been considered in [13] and [25] (and more recently in 
[27]). It would be interesting to apply these
ideas to the particular MHD models we have considered here.  
 
\vspace{4ex}
\begin{center}
{\bf\Large Acknowledgments}
\end{center}
We would like to thank Oleg Soloviev for useful discussions. The work of SS
was supported by the A. Onassis Public Benefit Foundation, and that of ST
by the Royal Society.

\vspace{2ex}
\begin{center}
\noin{\bf\Large References}
\end{center}
\vspace{2ex}
\begin{description}
\item{[1]} R.H. Kraichnan, `Inertial ranges in two-dimensional turbulence',
{\it Phys. Fluids } {\bf 10} (1967) 1417.
\item{[2]} A.M. Polyakov, Princeton Univ. Preprint PUPT-1341, hep-th/9209046.
\item{[3]} A.M. Polyakov {\it Nucl. Phys.} {\bf B396} (1993) 367.
\item{[4]} D.A. Lowe {\it Mod. Phys. Lett.} {\bf A8} (1993) 923.
\item{[5]} G. Falkovich and A. Hanany, `Spectra of Conformal Turbulence ',
Preprint WIS-92-88-PH, hep-th/9212015 (1992).
\item{[6]} Y. Matsuo {\it Mod. Phys. Lett.} {\bf A8} (1993) 619.
\item{[7]} H. Cateau, Y. Matsuo and M. Umeski,  `Predictions on two-dimensional
turbulence by conformal field theory ', Preprint UT-652, hep-th/9310056 (1993).
\item{[8]} B.K. Chung, S. Nam, Q-H. Park and H.J. Shin, {\it Phys. Lett.} {\bf
B309}
(1993) 58.
\item{[9]} B.K. Chung, S. Nam, Q-H. Park and H.J. Shin, {\it Phys. Lett.} {\bf
B317} (1993) 92.
\item{[10]} G.Falkovich, A.Hanany, {\it Phys. Rev. Lett. } {\bf 71 } (1993) 3454-3457,
hep-th/9301030
\item{[11]} M.A.I. Flohr {\it Nucl.Phys.} {\bf B482}
(1996) 567-578, hep-th/9606130.
\item{[12]} V. Gurarie, {\it Nucl.Phys.} {\bf B410} (1993) 535,
  hep-th/9303160
\item{[13]} J.S. Caux, I.I. Kogan, A.M. Tsvelik, {\it Nucl.Phys.} {\bf B466}
 (1996) 444.
\item{[14]} M.R.Rahimi Tabar, S. Rouhani, {\it The Alf'ven Effect and
 Conformal Field 
Theory }, hep-th/9507166; {\it A Logarithmic Conformal Field Theory
Solution For Two Dimensional Magnetohydrodynamics In Presence of The
Alf'ven Effect}, hep-th/9606143.
\item{[15]} A. Shafiekhani, M.R.Tabar, Logarithmic Operators in Conformal Field Theory
and the $ W_{\infty }$, hep-th/9604007
\item{[16]}  M. A.I. Flohr {\it Int.J.Mod.Phys.} {\bf A12} (1997)
1943-1958, hep-th/9605151
\item{[17]} L.C. Woods  `Principles of Magnetoplasma Dynamics ',
Oxford Science publs. (1987).
\item{[18]} G. Ferretti and Z. Yang, {\it Europhys. Lett.} {\bf 22}  (1993)
639.
\item{[19]} O. Coceal and S. Thomas, {\it Mod. Phys. Lett.}{\bf  A10} (1995)
\item{[20]} O. Coceal, W.A. Sabra and S. Thomas, {\it Europhys. Lett.}
{\bf 35} (5) (1996) 277.
\item{[21]} O. Coceal, W.A. Sabra and S.Thomas, {\it Europhys. Lett.}
{\bf 35} (5) (1996) 343.
\item{[22]} O. Coceal and S.Thomas, {\it Mod. Phys. Lett} {\bf A10} (1995)
2427.
\item{[23]} Ph. Brax,{\it  A Renormalisation Group Analysis of 2d Freely
    Decaying Magnetohydrodynamic Turbulence}, hep-th/9606156
{\it The Coulomb Gas Behaviour of Two Dimensional Turbulence}, 
hep-th/9505111
\item{[24]}  L. Moriconi, {\it 3-D Perturbations in Conformal Turbulence},
hep-th/9508040
\item{[25]} I.I. Kogan and N.E. Mavromatos, {\it Phys.Lett.} {\bf B375}
(1996) 111.
\item{[26]} I.I. Kogan and A. Lewis, {\it Nucl.Phys.} {\bf B509} (1998) 687.
\item{[27]} M.R. Rahimi Tabar, S. Rouhan {\it Zamalodchikov's C-Theorem and
The Logarithmic Conformal Field Theory}, hep-th/9707060.
\end{description}

\end{document}